\documentclass[%
reprint,
superscriptaddress,
 amsmath,amssymb,
 aps,
floatfix, 
]{revtex4-2}

\usepackage{graphicx}
\usepackage{bm}

\usepackage[dvipsnames]{xcolor}

\def\cm{{\,\textrm{cm}}}
\def\K{{\,\textrm{K}}}
\def\E{{\bm{E}}}
\def\eVAA{\, \textrm{eV/\AA}}
\def\eV{{\,\textrm{eV}}}

\begin{document}

\preprint{APS/123-QED}

\title{RASCBEC: RAman Spectroscopy Calculation via Born Effective Charge}


\author{Rui Zhang}
\affiliation{Department of Physics, Northeastern University, Boston MA 02115}
\affiliation{Department of Physics, University of Florida, Gainesville, Florida 32611}
\affiliation{Quantum Theory Project, University of Florida, Gainesville, Florida 32611}

\author{Jun Jiang}
\affiliation{Department of Physics, University of Florida, Gainesville, Florida 32611}
\affiliation{Quantum Theory Project, University of Florida, Gainesville, Florida 32611}

\author{Alec Mishkin}
\affiliation{Department of Physics, University of Florida, Gainesville, Florida 32611}
\affiliation{Quantum Theory Project, University of Florida, Gainesville, Florida 32611}

\author{James N. Fry}
\affiliation{Department of Physics, Northeastern University, Boston MA 02115}
\affiliation{Department of Physics, University of Florida, Gainesville, Florida 32611}

\author{Hai-Ping Cheng}%
\email{hping@ufl.edu}
\affiliation{Department of Physics, Northeastern University, Boston MA 02115}
\affiliation{Department of Physics, University of Florida, Gainesville, Florida 32611}
\affiliation{Quantum Theory Project, University of Florida, Gainesville, Florida 32611}

\date{\today}

\begin{abstract}

We advance the algorithm for ab initio calculations of Raman spectra for large systems via applying external electric field, and complement it by a code implementation we name RASCBEC. With the RASCBEC code, we have successfully benchmark crystalline materials and compute Raman spectra of large molecules, and amorphous oxides. Our results demonstrate a remarkable level of agreement with the results from other commonly used codes as well as the experimental data. 
The electric field approach for Raman spectra calculation is designed to overcome the computational challenges associated with the conventional approach, which requires the calculation of the macroscopic dielectric tensor at numerous molecular geometries. The key innovation in our approach lies in obtaining the first-order derivatives with respect to the external electric field directly from VASP (the Vienna Ab Initio Simulation Package), as the Born Effective Charge (BEC).
The RASCBEC code not only significantly reduces computational time, up to a factor of $N/8$, compared to the conventional approach, where $N$ is the total count of atoms within the simulation box. But also maintains the same level of accuracy, employing first-order numerical derivatives that avoid the numerical noise associated with algorithms requiring second-order derivatives, as seen in other electric field-based methods. This advantage makes RASCBEC particularly beneficial for large molecules and expansive amorphous systems.

\end{abstract}

\maketitle


\section{\label{sec:1}Introduction} 

Raman spectroscopy \cite{long1977raman} plays a pivotal role in elucidating the atomic-level structure of disordered systems, offering insights into the vibrational states of atoms. Thanks to enhanced performance and sensitivity, this technique has found extensive application in the characterization and analysis of diverse materials \cite{schrader2008infrared, colthup2012introduction, smith2019modern, li2016raman, vitol2009situ, baert2011raman, kovcivsova2016thiol}. 

The computational complexity associated with determining Raman intensity from a first-principles approach parallels that required for calculating vibrational modes \cite{PhysRevB.54.7830}.  To compute the derivative of the dipole moment with respect to the nuclear coordinates, the macroscopic dielectric tensor needs to be evaluated for a number of molecular geometries that is on the order of six times the number of atoms \cite{steele1972theory}. For large systems, the computational expense of a multitude of ab initio calculations is considerable. Several widely-used codes are available online to facilitate these calculations. Notable examples include ``vasp\_raman''  developed by Fonari and Stauffer in 2013 \cite{vasp_raman_py}, and ``Phonopy-Spectroscopy,'' created by Skelton et al. in 2017 \cite{skelton2017lattice}. These codes have undergone rigorous testing, proving their accuracy and reliability, and have been successfully applied to a wide array of materials.

To overcome the computational challenges typically associated with large systems, and thereby expand the applicability to a broad spectrum of polyatomic molecules, an alternative approach was introduced by Komornicki and McIver in 1979 \cite{komornicki1979efficient}. This approach predicts Raman activity from potential energy gradient derivatives with respect to an external electric field. Mathematically equivalent to the widely-used method, it offers a computational advantage as it requires the evaluation of the dipole moment and its derivatives only once at a single nuclear geometry. This is because these properties do not depend on the specific value of the applied field. Subsequent to the initial proposition and further development \cite{komornicki792, komornicki83} of this idea, Porzag and Pederson discussed its computational framework in 1996 \cite{PhysRevB.54.7830}. A more comprehensive version, including implementation within Quantum Espresso \cite{giannozzi2009quantum}, was provided by Umari1 and Pasquarello in 2005 \cite{umari2005infrared}. This method has demonstrated its effectiveness, especially in the case of various amorphous systems \cite{giacomazzi2005medium, giacomazzi2007vibrational, giacomazzi2009medium, khoo2010first, kilymis2019vibrational, giacomazzi2023infrared, giacomazzi2023infrared}. It has shown its capability to handle large models, with sizes of up to 200 atoms.

One of the notable advantages that distinguishes this method from other approaches aiming to accelerate Raman calculations \cite{rlarge1, collins2015energy, raghavachari2015accurate, doi:10.1021/ar500038z, Khire} is its relative ease of implementation and its exceptional efficiency in saving computational time, especially when dealing with amorphous materials. However, there are certain limitations to this method. First, conducting Self Consistent Field (SCF) calculations with a finite electric field can be problematic, as highlighted in previous research \cite{PhysRevB.54.7830}, adding a level of complexity to the method since reliable Raman intensities can only be obtained with well-converged wave functions. Second, the relative error of numerically determined polarizability derivatives is highly dependent on the magnitude of the electric field. This means that the electric field cannot be too large or too small, and this condition can be particularly limiting when studying materials with small band-gaps \cite{umari2005infrared}.

In this paper, we present a different implementation of this method within the Vienna Ab-initio Simulation Package (VASP) \cite{PhysRevB.54.11169, CMS_6_15}. What sets our approach apart from the previous setup \cite{umari2005infrared} is our method of obtaining the first-order derivatives with respect to the external electric field directly from VASP, as the Born Effective Charge (BEC). Instead of deriving forces and employing a 5-point formula for second-order numerical derivatives by the external electric field, we are able to utilize a 3-point formula for the first-order derivative of the BEC. This approach offers two key advantages. First, it enhances the efficiency and accuracy. The total number of self-consistent minimizations required for the electric-field-dependent energy functional is reduced from 25 to 8, which gains substantial time savings while maintaining the same degree of accuracy. Secondly, our approach widens the range of materials to which this method can be applied, particularly to materials with small band gaps. This enhancement is achieved by reducing the number of differencing points from 5 to 3, resulting in a more conservative range for the required electric field. Our tests of this scheme on various materials, including crystals, small band-gap large molecules, and amorphous oxides, demonstrate that our method provides results of comparable accuracy to the widely-used approach while reducing computational costs by a factor of $N/8$.

The paper is structured as follows: In Section \ref{sec:2}, we provide a comprehensive description of our method, outlining the scheme and deriving the formula for calculating Raman spectra from the Born effective charge. Section \ref{sec:3} is dedicated to the testing and validation of our implementation, which is performed using VASP. We begin by examining a crystal, GeO$_2$ and then extend our method to a small band-gap magnetic molecule consisting of 448 atoms and to a 350-atom sample of amorphous Ta$_2$O$_5$. In this section, we also delve into the comparison of our results with those obtained using other widely-used codes and with experimental data. Additionally, we explore the optimal strength of the applied electric field used in computing the electric field gradient derivatives. Section \ref{sec:4} encapsulates the final conclusions drawn from our research.

\section{\label{sec:2}The RASCBEC Method}

The conventional method calculates non-resonant first order Raman activity $I_{\mathrm{Raman},s}$ by the change in the polarisability tensor $\alpha_s$ along the mode eigenvectors $Q_{s}$ for each phonon mode $s$ within the Placzek approximation \cite{placzek1959rayleigh}, 
\begin{equation}
I_{\mathrm{Raman},s} = \operatorname{const} \, \left(\omega_{L}-\omega_{s}\right)^{4} \frac{\hbar}{2 \omega_{s}} \left|\hat{e}_{i} \, \frac{\partial \alpha_s}{\partial Q_{s}} \, \hat{e}_{j}\right|^{2} \left(1+\bar{n}_{s}\right) , 
\label{eqn:i}
\end{equation}
where 
$ \bar{n}_{s}=\left[\exp \left(\hbar \omega_{s} / k_{\mathrm{B}} T\right)-1\right]^{-1} $
is the mean occupation number of the phonon mode $s$, $\omega_L$ is the frequency of the incident light, $\omega_s$ is the frequency of phonon mode $s$, and $\hat{e}_i$ and $\hat{e}_j$ are the unit vectors of the electric field direction (polarization) for the scattered and the incident light.

The numerical computation of the polarizability tensor $\alpha_s$ is reformulated in terms of the macroscopic high-frequency dielectric constant $\varepsilon_s^{\infty}$ \cite{Mitroy_2010}, 
\begin{equation}
I_{\mathrm{Raman},s} \propto \left|\frac{\partial \alpha_s}{\partial Q_s}\right|^{2} \equiv \left|\frac{\partial \varepsilon_s^{\infty}}{\partial Q_s}\right|^{2} \approx \left|\frac{\Delta \varepsilon_s^{\infty}}{\Delta Q_s}\right|^{2} , 
\label{eqn:ra}
\end{equation}
and the Raman activity is computed using the central-difference scheme 
\begin{equation}
I_{\mathrm{Raman},s,ij}\propto \left|
\frac{\varepsilon_{s,ij}^{\infty}(+\delta)   - \varepsilon_{s, ij}^{\infty}(-\delta)} {2\delta}  \right|^{2}, 
\label{eqn:central}
\end{equation}
where $\varepsilon_{s,ij}^{\infty}(\pm \delta)$ are the components of the dielectric tensor $\varepsilon_{s}^{\infty}$ evaluated at positive and negative displacements $ \Delta Q_s \equiv 2\delta $ along the mode $s$. 

In a system containing $N$ atoms, computing its Raman intensities necessitates the calculation of a total of $6N$ dielectric tensors. This can be observed from Eqs.~(\ref{eqn:i}) to (\ref{eqn:central}). In these equations, $N$ atoms contribute to the generation of $3N$ phonon modes. For each phonon mode, calculating the dielectric tensor requires atomic displacements in both positive and negative directions. As $N$ increases, the computational time required for this method becomes significant.

To accelerate the process, the central idea is to avoid the computational burden associated with $N$ by rearranging the order of the derivative sequence of the polarizability tensor:
\begin{equation}
\frac{\partial {\alpha_{ij}}}{\partial Q_s} = \sum_{k} \sum_t \frac{\partial {\alpha_{ij}}}{\partial \xi_{kt}} \, \frac{\partial {\xi_{kt}}}{\partial Q_s} = \sum_k \sum_t \frac{\partial Z_{ikt} }{\partial E_j} \, \frac{\partial {\xi_{kt}}}{\partial Q_s} .
\label{eqn:aq}
\end{equation}
Here we utilized the Born Effective Charge (BEC), with elements $Z_{ij}$. $Z_{ij}$ is defined as the change of polarization in the $i$-direction $P_i$ with respect to the displacement $\xi_{t,j}$ of a given ion/atom $t$ in the $j$-direction, presented in Eq.~(\ref{eq:z1}), equivalently as the derivative of the Hellman-Feynman force $\bm{F}$ in the $i$-direction on ion/atom $t$ with respect to the external electric field $\E$ in the $j$-direction, shown in Eq.~(\ref{eq:z2}).
\begin{equation}
Z_{ij,t}=\frac{\partial P_i\,}{\partial \xi_{t,j}}
\label{eq:z1}
\end{equation}
\begin{equation}
Z_{t,ij}=\frac{\partial F_{t,i}}{\partial E_j}
\label{eq:z2}
\end{equation}
By using Eq.~(\ref{eqn:aq}) in Eq.~(\ref{eqn:ra}), the Raman activity is now calculated by taking the derivative of the BEC as presented in Eq.~(\ref{eqn:becr}),
\begin{equation}
I_{\mathrm{Raman},s} \propto \left| \frac{\partial \alpha_s}{\partial Q_s}\right|^{2} = \left|\sum_k \sum_t \frac{\partial Z_{ikt} }{\partial E_j} \, \frac{\partial {\xi_{kt}}}{\partial Q_s} \right|^{2}
\label{eqn:becr}
\end{equation}
The BEC is a built-in capability of VASP \cite{PhysRevB.54.11169, CMS_6_15} and $\partial {\xi_{kt}}/{\partial Q_s} $ is nothing but a coordinate transformation. The most complex aspect of this method is calculating the derivative of the BEC with respect to the electric field. 
Each atom's BEC is represented as a $3\times 3$ matrix, and the electric field is a $3\times 1$ vector. Therefore, the derivative becomes a $3\times 3 \times 3$ tensor, 
\begin{equation}
\frac{\partial Z_{ik}}{\partial E_j}=\frac{\partial^2 F_i} {\partial E_j \, \partial E_k} .
\label{eqn:delta}
\end{equation}
Thanks to the symmetry in $E_j$, $E_k$, this tensor possesses only 18 independent components.
By expanding the Hellman-Feynman force in a power series up to the third order in the electric field $\E$, we can express BEC as shown in Eq.~(\ref{eq:ep}) for applied electric fields $\E=(\Delta,0,0)$, $\bm{E}=(0,\Delta,0)$, and $\E=(0,0,\Delta)$ and symmetrically as in Eq.~(\ref{eq:en}) for applied electric fields $\E=(-\Delta,0,0)$, $\bm{E}=(0,-\Delta,0)$, and $\E=(0,0,-\Delta)$.
\begin{equation}
Z_{ij}=\frac{F_i (\E)-F_i (0)}{\Delta}=\frac{\partial F_i}{\partial E_j}
+\frac{1}{2}\frac{\partial^2F_i}{\partial E_j^2}\Delta+\frac{1}{6}\frac{\partial^3F_i}{\partial E_j^3}\Delta^2
\label{eq:ep}
\end{equation}
\begin{equation}
Z_{i,-j}=\frac{\partial F_i}{\partial E_j}
-\frac{1}{2}\frac{\partial^2F_i}{\partial E_j^2}\Delta+\frac{1}{6}\frac{\partial^3F_i}{\partial E_j^3}{\Delta}^2
\label{eq:en}
\end{equation}
where the index $-j$ denotes applying  the electric field in the  negative-$j$ direction.
Then some of the components of the derivative tensor can be immediately computed by Eq.~(\ref{eqn:ijj}).
\begin{equation}
\frac{\partial Z_{ij}}{\partial E_j}=\frac{\partial^2F_i}{\partial E_j^2}=\frac{Z_{ij}-Z_{i,-j}}{\Delta}
\label{eqn:ijj}
\end{equation}
It is important to note that in Eq.~(\ref{eqn:ijj}), the second index of $Z_{ij}$ must be the same as the index of $E_j$ because each derivative considers changes in the electric field along the same direction. Eq.~(\ref{eqn:ijj}) covers 9 out of the 27 components required.
Due to the sequential application of an electric field in the $x$, $y$, and $z$-directions in VASP, it is not possible to directly apply an electric field in a diagonal direction. To calculate off-diagonal terms, we employ a rotation of the lattice, for example rotating it by an angle of $-\pi/4$ in the $x$-$y$ plane then calculating the BEC with this rotated structure and a new electric field, $\E'=(\sqrt{2}\Delta,0,0)$, where $x'$ represents the horizontal direction in the rotated frame and corresponds to a diagonal direction in the $x$-$y$ plane relative to the original lattice. This rotation allows us to obtain the off-diagonal derivative terms. The details will be as follows:
For example, to get the off-diagonal term $\frac{\partial Z_{zx}}{\partial E_y}$ we consider $Z_{zx'}$ which is obtained by applying electric field $\E'=(\Delta,\Delta,0)$. In Eq.~(\ref{eqn:zx}) $Z_{zx'}$ is expressed in the coordinate system of lattice vectors. 
\begin{equation}
Z_{zx'}=\frac{F_z(\Delta,\Delta,0)-F_z(0,0,0)}{\sqrt{2}\Delta}
\label{eqn:zx}
\end{equation}
By expanding $F_z(\Delta, \Delta, 0)$ to second order in $\Delta$, the expression becomes Eq.~(\ref{eq:fz1}) (with unnecessary zero arguments omitted). A similar expression can be obtained in Eq.~(\ref{eq:fz2}) for $Z_{z,-x'}$ by applying electric field $\E'=(-\Delta,-\Delta,0)$.
\begin{eqnarray}
&&F_z(\Delta,\Delta,0) = F_z(0) + 
\Bigl(\frac{\partial F_z}{\partial E_x}  + \frac{\partial F_z}{\partial E_y}\Bigr) \, \Delta
\nonumber\\
&&\qquad  + \Bigl( \frac{1}{2}\frac{\partial^2 F_z}{\partial E_x^2} +\frac{1}{2}\frac{\partial^2 F_z}{\partial E_y^2} 
 +\frac{\partial^2 F_z}{\partial E_x\partial E_y} \Bigr) \Delta^2+ O(\Delta^3)
\nonumber\\
&&\quad = F_z(0)+(Z_{zx}+Z_{zy})\Delta
  +\frac{\partial Z_{zx}}{\partial E_y}\Delta^2+O(\Delta^3)
\\
\label{eq:fz1}
%
\nonumber\\
\noalign{\bigskip}
&&F_z(-\Delta,-\Delta,0) = F_z(0) \nonumber\\
&&\quad - (Z_{z,-x}+Z_{z,-y}) \, \Delta 
+\frac{\partial Z_{zx}}{\partial E_y} \, \Delta^2 + O(\Delta^3)
\label{eq:fz2}
\end{eqnarray}
\begin{eqnarray}
&&Z_{zx'}-Z_{z,-x'} \nonumber\\
&&\quad = \frac{\left[ F_z(\Delta,\Delta,0)+F_z(-\Delta,-\Delta,0) 
 - 2F_z(0) \right]}{\sqrt{2}\Delta } \\
&&\quad=\sqrt{2}\,\frac{\partial Z_{zx}}{\partial E_y} \, \Delta
 + \frac{1}{\sqrt{2}} \, (Z_{zx}-Z_{z,-x} +  Z_{zy}-Z_{z,-y}) . \nonumber 
\label{eqn:zxzx}
\end{eqnarray}
By subtracting $Z_{z,-x'}$ from $Z_{z,x'}$, the derivative ${\partial Z_{zx}}/{\partial E_y}$ can be determined to be
\begin{eqnarray}
\frac{\partial Z_{zx}}{\partial E_y}=\frac{\partial Z_{zy}}{\partial E_x}
&=& [\sqrt{2}(Z_{zx'}-Z_{z,-x'}) - Z_{zx} \nonumber \\
&& +Z_{z,-x} - Z_{zy} + Z_{z,-y}] / 2\Delta .
\label{eq:zzx}
\end{eqnarray}
We can follow the same procedure for forces in the other two directions, $F_{x'} = (F_x + F_y) / \sqrt{2}$ and $F_{y'} = -(F_x - F_y) / \sqrt{2}$, allowing us to calculate four more elements in the derivative tensor while taking symmetry into account, 
\begin{eqnarray}
\frac{\partial Z_{xx}}{\partial E_y}&&=\frac{\partial Z_{xy}}{\partial E_x}
=(Z_{x'x'}-Z_{x',-x'}-Z_{y'x'} +Z_{y',-x'} \nonumber\\
&&\qquad -Z_{xx}+Z_{x,-x} -Z_{xy}+Z_{x,-y})/2\Delta \\
\noalign{\medskip}
\frac{\partial Z_{yy}}{\partial E_x}&&=\frac{\partial Z_{yx}}{\partial E_y} 
= (Z_{x'x'}-Z_{x',-x'}+Z_{y'x'} - Z_{y',-x'} \nonumber\\
&&\qquad  -Z_{yx} + Z_{y,-x} - Z_{yy }+ Z_{y,-y})/2\Delta .
\label{eq:zxxzyx}
\end{eqnarray}
By performing similar rotations in the $xz$ and $yz$ planes, we can obtain the remaining 12 components of the derivative of BEC. Now that all elements of the derivative tensor in Eq.~(\ref{eqn:delta}) have been obtained, we can compute the Raman intensity tensor once the BECs are ready. BECs along the $x$, $y$, and $z$-directions can be computed in a single calculation with the unrotated system to save computation time. The same applies to BECs along the $-x$, $-y$, and $-z$-directions. The calculations using the rotated system yield BECs along the $xy$, $yz$, and $xz$-directions, both in the positive and negative directions. In total, the calculations yield 8 BECs. 

Our method offers several advantages: Firstly, in comparison to widely-used codes, our method requires only 8 Born Effective Charge (BEC) calculations, as opposed to the $6N$ dielectric tensor calculations, resulting in a significant reduction in computational time. Secondly, unlike previous setups aimed at accelerating Raman calculations \cite{umari2005infrared}, which involve 25 force calculations and utilize a 5-point formula for second-order numerical derivatives of forces with respect to the external electric field, our method directly computes 8 BECs from VASP and employs a 3-point formula for the first-order derivative of BEC. This approach substantially improves both efficiency and accuracy. Finally, our approach widens the range of materials to which this method can be applied, particularly those with small band gaps. This expansion is achieved by reducing the number of differencing points from 5 to 3, enabling a more conservative range for the required electric field. Consequently, our method can be applied to a broader variety of materials.

\section{\label{sec:3}Validation of the RASCBEC method}

In this section, we will assess and validate our method through the examination of three different systems: a rutile-type GeO$_2$ crystal, an amorphous sample of Ta$_2$O$_5$ consisting of 350 atoms, and a large magnetic molecule composed of 448 atoms. 
We start by using the rutile-type GeO$_2$ with a six-atom unit cell as the first test sample to validate RASCBEC. The simulation data for modelling rutile-type GeO$_2$ (mp-470) is obtained from the Materials Project \cite{osti_1208347}. Our approach is implemented through RASCBEC. To validate our method, we have chosen to compare with the well-established vasp\_raman code \cite{vasp_raman_py}, which utilizes the conventional method and is well-suited for small crystal systems. All phonon calculations, dielectric tensor calculations, and BEC calculations were conducted using the Vienna Ab-initio Simulation Package (VASP) \cite{PhysRevB.54.11169, CMS_6_15} with the generalized gradient approximation (GGA) Perdew-Burke-Ernzerhof (PBE) functional on AMD EPYC 7702 processors. A cutoff of $520 \eV$ is used to converge the energy to less than $1e-7 \eV$.

\begin{figure}
\centering
    \includegraphics[width=0.4\textwidth]{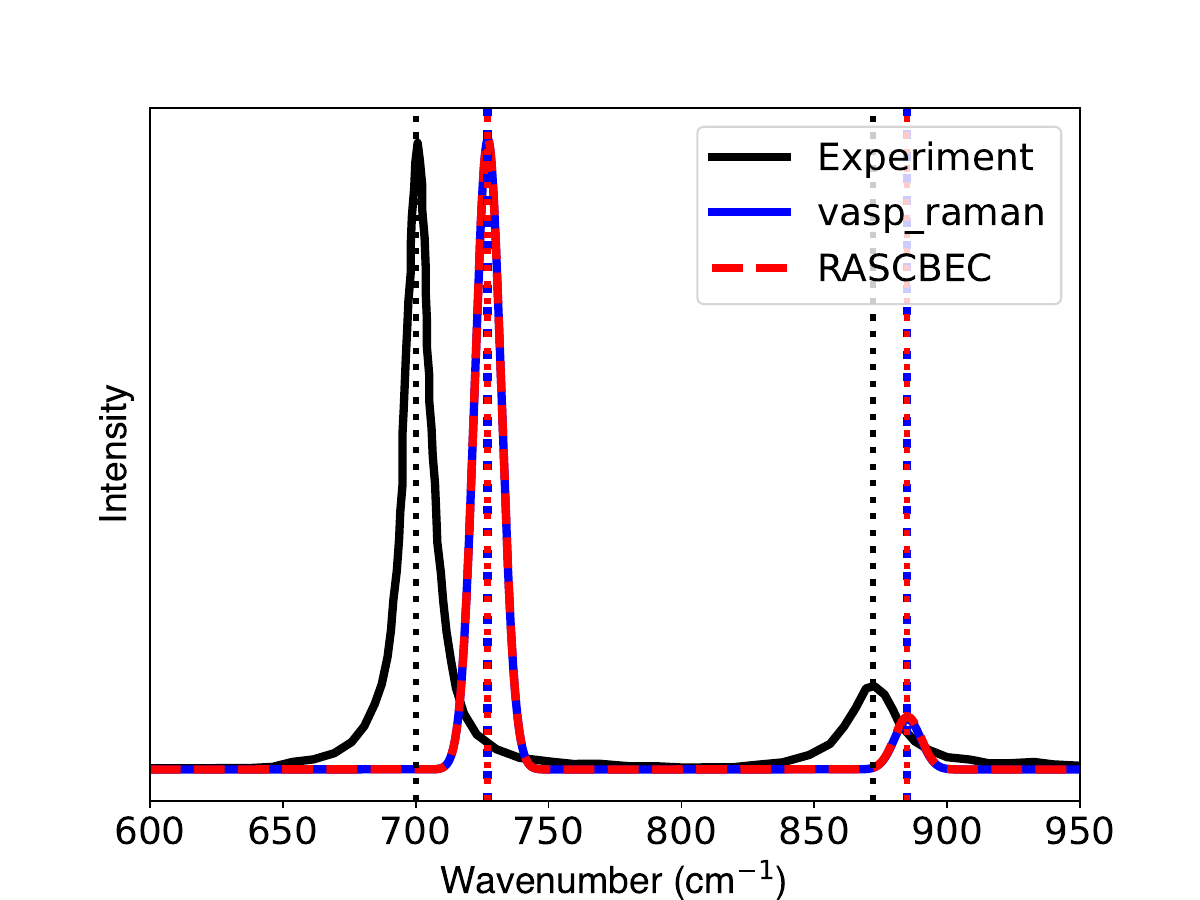}
    \caption{Experimental and calculated Raman spectra of a rutile-type GeO$_2$ crystal using vasp\_raman and RASCBEC.}
    \label{fig:g6}
\end{figure}

The calculated Raman spectra using RASCBEC and vasp\_raman \cite{vasp_raman_py} are illustrated in Figure~\ref{fig:g6}, with the experimental Raman spectrum of rutile GeO$_2$ measured at $298 \K$ \cite{mernagh1997temperature} also provided for reference. Both codes produce Raman spectra featuring two prominent peaks, which align with the experimental observations with slight shifts in the peak positions: the experimental measurements identify two peaks at $700 \cm^{-1}$ and $872 \cm^{-1}$, while the calculations place the peaks at $727 \cm^{-1}$ and $885 \cm^{-1}$, as indicated by the dotted lines in Figure~\ref{fig:g6}. It is worth noting that RASCBEC and vasp\_raman yield nearly identical results, with slight differences in peak heights. This demonstrates the validation and accuracy of RASCBEC for this particular system.

\begin{figure}
\centering
    \includegraphics[width=0.4\textwidth]{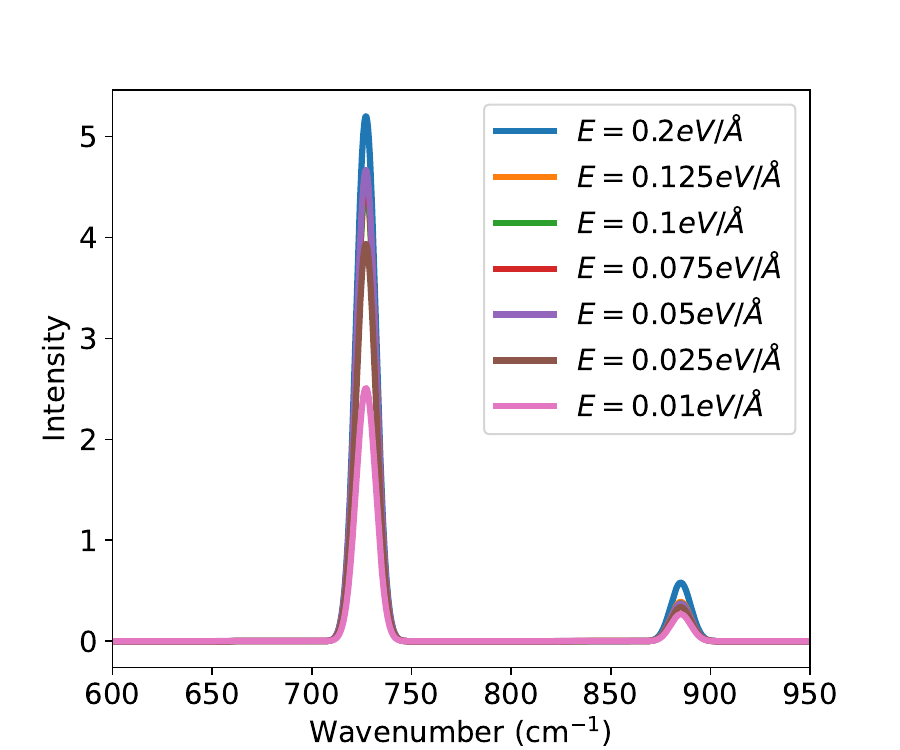}
    \caption{Raman spectra of a rutile-type GeO$_2$ crystal calculated using RASCBEC with different electric fields applied.}
    \label{fig:ecom}
\end{figure}

\begin{figure}
\centering
    \includegraphics[width=0.4\textwidth]{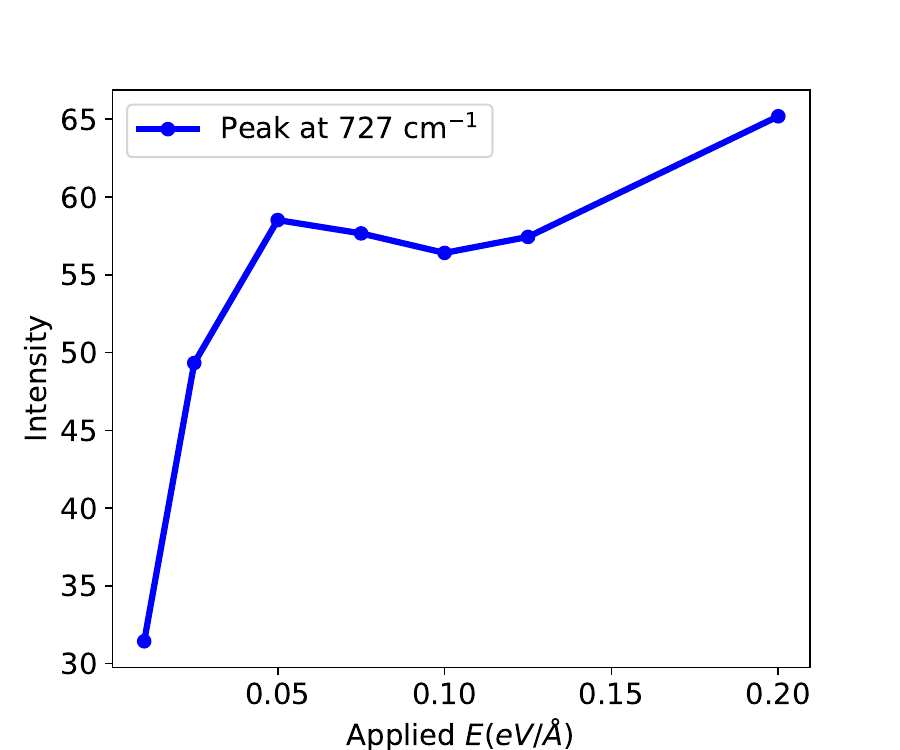}
    \caption{First peak height variation with electric field.}
    \label{fig:p1}
\end{figure}

\begin{figure}
\centering
    \includegraphics[width=0.4\textwidth]{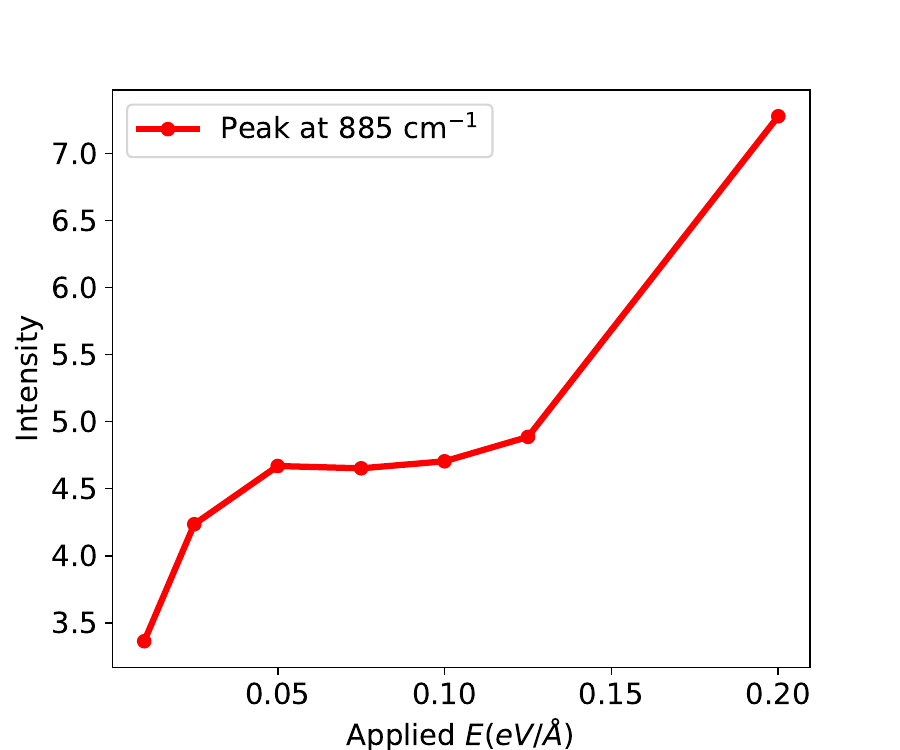}
    \caption{Second peak height variation with electric field.}
    \label{fig:p2}
\end{figure}

We should note that the accuracy of the Raman spectra calculations with RASCBEC can be influenced by the magnitude of the applied electric field ($\Delta$) used to compute the electric field gradient derivatives. If the field is too small, the errors may be dominated by DFT convergence and round-off errors. Conversely, if the field is too large, higher-order finite difference errors can become significant. A test was conducted with varying electric field strengths, ranging from $0.025 \eVAA$ to $0.2 \eVAA$, and the resulting Raman spectra are shown in Figure~\ref{fig:ecom}. It is evident that the  intensities of the Raman peaks change with the strength of the electric field. Figure~\ref{fig:p1} and Figure~\ref{fig:p2} illustrate how the intensity varies with the electric field strength. There is a relatively flat region between $0.05 \eVAA$ and $0.125 \eVAA$, indicating that within this range, the computational results are stable. The results presented in Figure~\ref{fig:g6} used an electric field of $E = 0.1 \eVAA$. This choice falls within the stable region where computational results are robust.

Timing information for both the RASCBEC and vasp\_raman is presented in Table ~\ref{tab:g6}. It is worth noting that the estimated time for calculating $\epsilon$ or $Z$ given in the table should be considered an upper limit. This is because all BEC calculations performed after the first one are faster. After the initial rotation, only one direction of the electric field needs to be applied, as opposed to the three directions during the first calculation. 

\begin{table*}
\caption{\label{tab:table6}Timing information of RASCBEC and vasp\_raman tested on rutile-type GeO$_2$ crystal (6 atoms)}
\begin{ruledtabular}
\begin{tabular}{cccc}
  $N=6$               & Number of modes or Z  & Average time of getting 1 $\epsilon$ or $Z$ & Total time        \\
  \hline
  Dielectric tensor ($\epsilon$) & $6(N-1)=30$      & ~1,700 sec (1 cpu)  & ~51,000 sec \\
  BEC ($Z$)                 & 8    & 24,886 sec (1 cpu)  & 200,000 sec \\
\end{tabular}
\end{ruledtabular}
\label{tab:g6}
\end{table*}

\begin{table*}
\caption{\label{tab:table448}Timing information of RASCBEC and Phonopy-Spectrosscopy tested on a magnetic molecule (448 atoms)}
\begin{ruledtabular}
\begin{tabular}{cccc}
  $N=448$              & Number of modes or Z  & Average time of getting 1 $\epsilon$ or $Z$ & Total time        \\
  \hline
  Dielectric tensor ($\epsilon$)  & $6(N-1)=2682$         & ~20,450 sec (32 cpu)     & 54,846,900 sec \\
  BEC ($Z$)           & 8        & 122,860 sec (32 cpu)      & ~~~982,880 sec \\
\end{tabular}
\end{ruledtabular}
\label{tab:m}
\end{table*}

The comparison between RASCBEC and commonly-used codes in terms of computational efficiency is based on the number of dielectric tensor and BEC calculations required. To put this into perspective, for a system of $N$ atoms, the conventional method will need to generate $6N$ dielectric tensors. On the other hand, RASCBEC needs only 8 BEC calculations, each with an electric field applied, and each of these calculations takes approximately 18 times longer than one without an electric field. As a result, the time ratio between the conventional method and RASCBEC is approximately $N/8$. This implies that when $N$ is greater than 8, RASCBEC becomes significantly more computationally efficient than the conventional method.

\begin{figure}
\centering
    \includegraphics[width=0.5\textwidth]{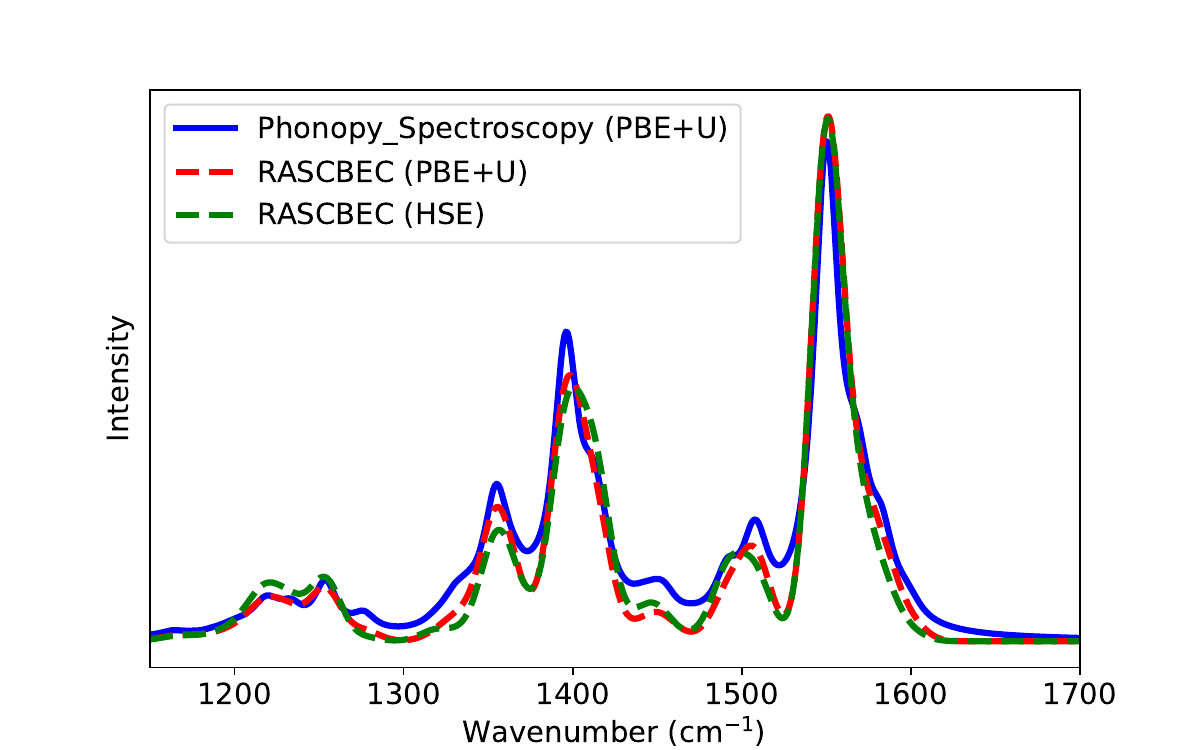}
\caption{Calculated Raman spectra of the Fe\_SCO magnetic molecule using Phonopy-Spectroscopy and RASCBEC}
    \label{fig:fer}
\end{figure}

To highlight the advantages of our RASCBEC code for handling large systems, we selected the spin-crossover (SCO) magnetic molecule [Fe($t$Bu$_2$qsal)$_2$], composed of 448 atoms, which has gained attention in recent research \cite{doi:10.1021/jacs.1c04598}, and computed the Raman spectra of this molecule's low spin state using both RASCBEC and a well-established Phonopy-Spectroscopy code \cite{skelton2017lattice}.

As depicted in Figure~\ref{fig:fer}, the red and green curves represent the Raman spectra calculated using RASCBEC. The red curve is obtained using PBE+$U$ to compute the BEC, while the green curve employs a hybrid functional. In both cases, an electric field of $E = 0.05 \eVAA$ is applied, and the intensities of these two spectra are normalized using the same factor for comparison with the Raman spectrum calculated through Phonopy-Spectroscopy, which is presented by the blue dashed curve. It is evident that all three curves share common peak features, including the phonon frequencies corresponding to each peak and the relative intensities among the peaks. However, there are still discernible differences at specific frequencies. These differences likely stem from the approximations made when computing derivatives numerically, whether the derivative of the dielectric tensor in the conventional method or the derivative of the BEC in RASCBEC. When calculating derivatives, all numerical methods involve some degree of approximation. In the context of solids, these approximations tend to be more reliable than in molecules. In molecules, certain vibrational modes can exhibit large vibrations, and anharmonic effects may become more pronounced. Nonetheless, even in the presence of strong anharmonicity, one can select an electric field strength such that the response to the electric field remains approximately linear. This linearity is evident in Figures~\ref{fig:p1} and \ref{fig:p2}, where the effects of the electric field are more pronounced as the field strength increases.

\begin{figure}
\centering
    \includegraphics[width=0.5\textwidth]{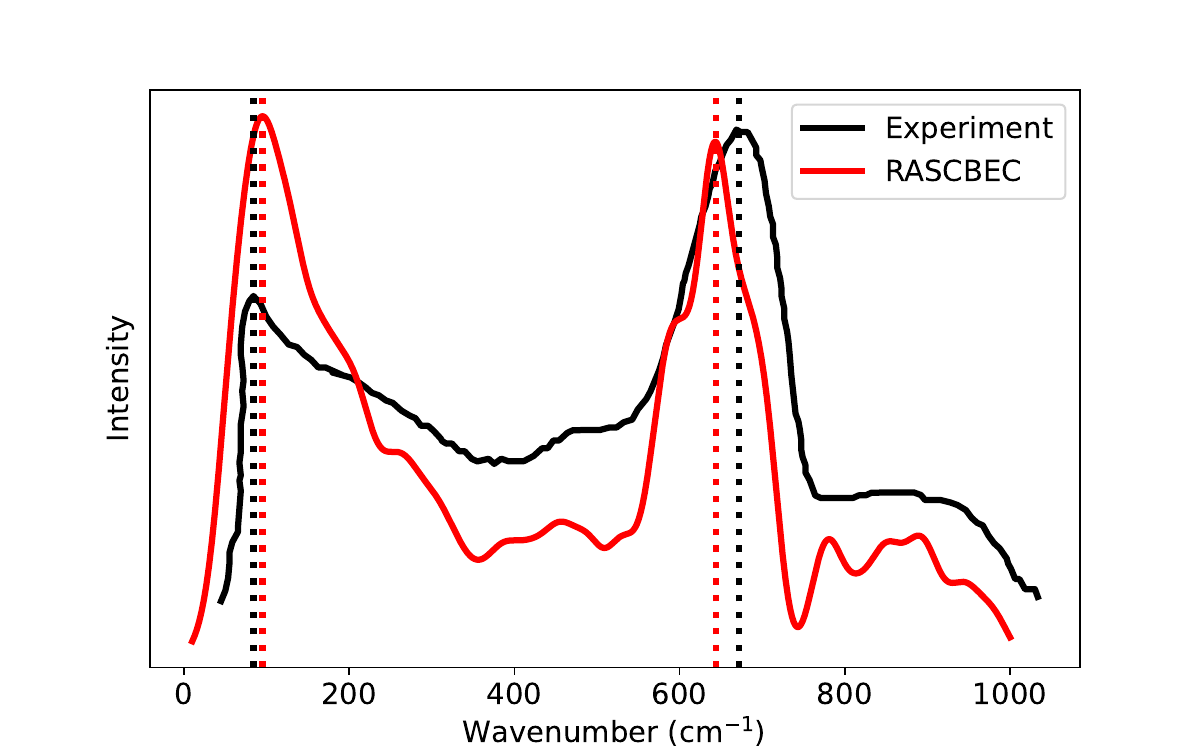}
    \caption{Calculated Raman spectra of amorphous Ta$_2$O$_5$ using RASCBEC compared with experimental data \cite{doi:10.1002/jrs.3142}.}
    \label{fig:taa}
\end{figure}

The timing information for both RASCBEC and Phonopy-Spectroscopy applied on this magnetic molecule is presented in Table ~\ref{tab:m}. All computations were conducted on AMD EPYC 7702 cores, utilizing the Vienna Ab-initio Simulation Package (VASP) \cite{PhysRevB.54.11169, CMS_6_15} for first-principle calculations. In terms of total computation cost, the ratio between the conventional method and RASCBEC is $54846900/982880 = 55.8$, while the predicted ratio is $N/8 = 448/8 = 56$. This demonstrates that RASCBEC is capable of saving a factor of 56 in computation time for a 448-atom molecule while producing Raman spectra with the same level of accuracy.

Finally we test RASCBEC on large and complex systems: amorphous materials. Handling large and disordered systems with thousands of first-principle calculations can indeed be a challenging task, given the significant computational resources required. Fortunately, RASCBEC is a valuable tool that can help manage and streamline such complex calculations. We choose amorphous Ta$_2$O$_5$ as our test material. An amorphous sample containing 350 atoms is prepared using a Molecular Dynamics (MD) simulation with the melt-quench method. The MD simulation was conducted using the classical MD simulation code LAMMPS \cite{PLIMPTON19951} with a classical potential formulated by Trinastic et al. \cite{doi:10.1063/1.4825197}. The amorphous Ta$_2$O$_5$ sample was enclosed in a cubic box with a box size of $16.43 \, \textrm{\AA} $. The preparation process involved heating the sample to a high temperature of 6000 K and then gradually cooling it down to $300 \K$. This cooling process was performed at a rate of $0.33 \K$ per picosecond. After the cooling phase, we performed structural relaxation using Density Functional Theory (DFT). The slow cooling rate employed in our computational approach was chosen to ensure that the resulting amorphous Ta$_2$O$_5$ structure closely mimics real-world conditions to the best of our abilities. This allowed us to create an amorphous Ta$_2$O$_5$ structure suitable for subsequent Raman spectra calculations.

The calculated and experimental Raman spectra \cite{https://doi.org/10.1002/jrs.3142} are presented in Figure~\ref{fig:taa}. The black curve represents the experimental measurement, while the red curve corresponds to the Raman spectrum calculated using RASCBEC. Dotted lines indicate the positions of the peaks. It is evident that both the experimental and calculated spectra exhibit two main peaks, and these peaks are closely aligned. In the experiment, the first and second peaks are observed at $84 \cm^{-1}$ and $672 \cm^{-1}$, while in the calculation, they are found at $95 \cm^{-1}$ and $644 \cm^{-1}$. When using RASBEC for Raman calculations of amorphous Ta$_2$O$_5$, the computation time for a single BEC on an AMD EPYC 7702 processor is approximately $65,540$ seconds using 256 CPU cores. Therefore, the total computational cost is $524,320$ seconds. To provide context on the conventional method for amorphous materials, we attempted one dielectric tensor calculation, which took around $52,170$ seconds using 256 CPU cores. Extrapolating this to the 2094 such calculations required, the total computation time would be approximately $109,243,980$ seconds. This demonstrates the significant advantage in efficiency when employing RASCBEC for Raman spectra calculations.

\section{\label{sec:4}Conclusion}

In conclusion, in this paper we present a significant development of the method to calculate Raman spectra using the finite difference gradient with respect to an external electric field. We use this derivative, first-order in nature, of the Born effective charge to replace the sum over phonon modes of polarizability derivatives along mode eigenvectors used in the conventional method.
We call the new method the RASBEC.
Since for large systems there is a large number of terms in the polarizability sum, this allows us to expedite first-principles Raman calculations and probe large systems in an economical way. This method saves computation time for large systems by a factor of $N/8$, where $N$ is the number of atoms in the simulation box.
Tests of the RASBEC method on an atomic crystal, a large magnetic molecule, and a large amorphous system verify its ability to reproduce the results of the conventional method and its significant reduction in computational cost. Because of the connection with VASP, we foresee a great impact on the condensed matter and materials physics community.
Applying this method to calculate Raman activity for even larger and more complex systems is possible, and we will explore more directions in the future.


\begin{acknowledgments}

This work is supported by the LIGO coating center funded by the U.S. National Science Foundation through grants PHY-2011776, PHY-2011770 and PHY-2309291. Computations were performed using the utilities of the National Energy Research Scientific Computing Center (NERSC) and the University of Florida Research Computing HiPerGator.

During the preparation of this work the authors used ChatGPT 3.5 in order to improve language and readability. After using this tool/service, the authors reviewed and edited the content as needed and take full responsibility for the content of the publication.
\end{acknowledgments}


\bibliography{apssamp}

\begin{thebibliography}{39}%
\makeatletter
\providecommand \@ifxundefined [1]{%
 \@ifx{#1\undefined}
}%
\providecommand \@ifnum [1]{%
 \ifnum #1\expandafter \@firstoftwo
 \else \expandafter \@secondoftwo
 \fi
}%
\providecommand \@ifx [1]{%
 \ifx #1\expandafter \@firstoftwo
 \else \expandafter \@secondoftwo
 \fi
}%
\providecommand \natexlab [1]{#1}%
\providecommand \enquote  [1]{``#1''}%
\providecommand \bibnamefont  [1]{#1}%
\providecommand \bibfnamefont [1]{#1}%
\providecommand \citenamefont [1]{#1}%
\providecommand \href@noop [0]{\@secondoftwo}%
\providecommand \href [0]{\begingroup \@sanitize@url \@href}%
\providecommand \@href[1]{\@@startlink{#1}\@@href}%
\providecommand \@@href[1]{\endgroup#1\@@endlink}%
\providecommand \@sanitize@url [0]{\catcode `\\12\catcode `\$12\catcode
  `\&12\catcode `\#12\catcode `\^12\catcode `\_12\catcode `\%12\relax}%
\providecommand \@@startlink[1]{}%
\providecommand \@@endlink[0]{}%
\providecommand \url  [0]{\begingroup\@sanitize@url \@url }%
\providecommand \@url [1]{\endgroup\@href {#1}{\urlprefix }}%
\providecommand \urlprefix  [0]{URL }%
\providecommand \Eprint [0]{\href }%
\providecommand \doibase [0]{https://doi.org/}%
\providecommand \selectlanguage [0]{\@gobble}%
\providecommand \bibinfo  [0]{\@secondoftwo}%
\providecommand \bibfield  [0]{\@secondoftwo}%
\providecommand \translation [1]{[#1]}%
\providecommand \BibitemOpen [0]{}%
\providecommand \bibitemStop [0]{}%
\providecommand \bibitemNoStop [0]{.\EOS\space}%
\providecommand \EOS [0]{\spacefactor3000\relax}%
\providecommand \BibitemShut  [1]{\csname bibitem#1\endcsname}%
\let\auto@bib@innerbib\@empty
\bibitem [{\citenamefont {Long}(1977)}]{long1977raman}%
  \BibitemOpen
  \bibfield  {author} {\bibinfo {author} {\bibfnamefont {D.~A.}\ \bibnamefont
  {Long}},\ }\bibfield  {title} {\bibinfo {title} {Raman spectroscopy},\
  }\href@noop {} {\bibfield  {journal} {\bibinfo  {journal} {New York}\
  }\textbf {\bibinfo {volume} {1}} (\bibinfo {year} {1977})}\BibitemShut
  {NoStop}%
\bibitem [{\citenamefont {Schrader}(2008)}]{schrader2008infrared}%
  \BibitemOpen
  \bibfield  {author} {\bibinfo {author} {\bibfnamefont {B.}~\bibnamefont
  {Schrader}},\ }\href@noop {} {\emph {\bibinfo {title} {Infrared and Raman
  spectroscopy: methods and applications}}}\ (\bibinfo  {publisher} {John Wiley
  \& Sons},\ \bibinfo {year} {2008})\BibitemShut {NoStop}%
\bibitem [{\citenamefont {Colthup}(2012)}]{colthup2012introduction}%
  \BibitemOpen
  \bibfield  {author} {\bibinfo {author} {\bibfnamefont {N.}~\bibnamefont
  {Colthup}},\ }\href@noop {} {\emph {\bibinfo {title} {Introduction to
  infrared and Raman spectroscopy}}}\ (\bibinfo  {publisher} {Elsevier},\
  \bibinfo {year} {2012})\BibitemShut {NoStop}%
\bibitem [{\citenamefont {Smith}\ and\ \citenamefont
  {Dent}(2019)}]{smith2019modern}%
  \BibitemOpen
  \bibfield  {author} {\bibinfo {author} {\bibfnamefont {E.}~\bibnamefont
  {Smith}}\ and\ \bibinfo {author} {\bibfnamefont {G.}~\bibnamefont {Dent}},\
  }\href@noop {} {\emph {\bibinfo {title} {Modern Raman spectroscopy: a
  practical approach}}}\ (\bibinfo  {publisher} {John Wiley \& Sons},\ \bibinfo
  {year} {2019})\BibitemShut {NoStop}%
\bibitem [{\citenamefont {Li}\ \emph {et~al.}(2016)\citenamefont {Li},
  \citenamefont {Stanwix}, \citenamefont {Aman}, \citenamefont {Johns},
  \citenamefont {May},\ and\ \citenamefont {Wang}}]{li2016raman}%
  \BibitemOpen
  \bibfield  {author} {\bibinfo {author} {\bibfnamefont {H.}~\bibnamefont
  {Li}}, \bibinfo {author} {\bibfnamefont {P.}~\bibnamefont {Stanwix}},
  \bibinfo {author} {\bibfnamefont {Z.}~\bibnamefont {Aman}}, \bibinfo {author}
  {\bibfnamefont {M.}~\bibnamefont {Johns}}, \bibinfo {author} {\bibfnamefont
  {E.}~\bibnamefont {May}},\ and\ \bibinfo {author} {\bibfnamefont
  {L.}~\bibnamefont {Wang}},\ }\bibfield  {title} {\bibinfo {title} {Raman
  spectroscopic studies of clathrate hydrate formation in the presence of
  hydrophobized particles},\ }\href@noop {} {\bibfield  {journal} {\bibinfo
  {journal} {The Journal of Physical Chemistry A}\ }\textbf {\bibinfo {volume}
  {120}},\ \bibinfo {pages} {417} (\bibinfo {year} {2016})}\BibitemShut
  {NoStop}%
\bibitem [{\citenamefont {Vitol}\ \emph {et~al.}(2009)\citenamefont {Vitol},
  \citenamefont {Orynbayeva}, \citenamefont {Bouchard}, \citenamefont
  {Azizkhan-Clifford}, \citenamefont {Friedman},\ and\ \citenamefont
  {Gogotsi}}]{vitol2009situ}%
  \BibitemOpen
  \bibfield  {author} {\bibinfo {author} {\bibfnamefont {E.~A.}\ \bibnamefont
  {Vitol}}, \bibinfo {author} {\bibfnamefont {Z.}~\bibnamefont {Orynbayeva}},
  \bibinfo {author} {\bibfnamefont {M.~J.}\ \bibnamefont {Bouchard}}, \bibinfo
  {author} {\bibfnamefont {J.}~\bibnamefont {Azizkhan-Clifford}}, \bibinfo
  {author} {\bibfnamefont {G.}~\bibnamefont {Friedman}},\ and\ \bibinfo
  {author} {\bibfnamefont {Y.}~\bibnamefont {Gogotsi}},\ }\bibfield  {title}
  {\bibinfo {title} {In situ intracellular spectroscopy with surface enhanced
  raman spectroscopy (sers)-enabled nanopipettes},\ }\href@noop {} {\bibfield
  {journal} {\bibinfo  {journal} {ACS nano}\ }\textbf {\bibinfo {volume} {3}},\
  \bibinfo {pages} {3529} (\bibinfo {year} {2009})}\BibitemShut {NoStop}%
\bibitem [{\citenamefont {Baert}\ \emph {et~al.}(2011)\citenamefont {Baert},
  \citenamefont {Meulebroeck}, \citenamefont {Wouters}, \citenamefont {Ceglia},
  \citenamefont {Nys}, \citenamefont {Thienpont},\ and\ \citenamefont
  {Terryn}}]{baert2011raman}%
  \BibitemOpen
  \bibfield  {author} {\bibinfo {author} {\bibfnamefont {K.}~\bibnamefont
  {Baert}}, \bibinfo {author} {\bibfnamefont {W.}~\bibnamefont {Meulebroeck}},
  \bibinfo {author} {\bibfnamefont {H.}~\bibnamefont {Wouters}}, \bibinfo
  {author} {\bibfnamefont {A.}~\bibnamefont {Ceglia}}, \bibinfo {author}
  {\bibfnamefont {K.}~\bibnamefont {Nys}}, \bibinfo {author} {\bibfnamefont
  {H.}~\bibnamefont {Thienpont}},\ and\ \bibinfo {author} {\bibfnamefont
  {H.}~\bibnamefont {Terryn}},\ }\bibfield  {title} {\bibinfo {title} {Raman
  spectroscopy as a rapid screening method for ancient plain window glass},\
  }\href@noop {} {\bibfield  {journal} {\bibinfo  {journal} {Journal of Raman
  Spectroscopy}\ }\textbf {\bibinfo {volume} {42}},\ \bibinfo {pages} {1055}
  (\bibinfo {year} {2011})}\BibitemShut {NoStop}%
\bibitem [{\citenamefont {Ko{\v{c}}i{\v{s}}ov{\'a}}\ \emph
  {et~al.}(2016)\citenamefont {Ko{\v{c}}i{\v{s}}ov{\'a}}, \citenamefont
  {Proch{\'a}zka},\ and\ \citenamefont
  {{\v{S}}{\'\i}pov{\'a}}}]{kovcivsova2016thiol}%
  \BibitemOpen
  \bibfield  {author} {\bibinfo {author} {\bibfnamefont {E.}~\bibnamefont
  {Ko{\v{c}}i{\v{s}}ov{\'a}}}, \bibinfo {author} {\bibfnamefont
  {M.}~\bibnamefont {Proch{\'a}zka}},\ and\ \bibinfo {author} {\bibfnamefont
  {H.}~\bibnamefont {{\v{S}}{\'\i}pov{\'a}}},\ }\bibfield  {title} {\bibinfo
  {title} {Thiol-modified gold-coated glass as an efficient hydrophobic
  substrate for drop coating deposition raman (dcdr) technique},\ }\href@noop
  {} {\bibfield  {journal} {\bibinfo  {journal} {Journal of Raman
  Spectroscopy}\ }\textbf {\bibinfo {volume} {47}},\ \bibinfo {pages} {1394}
  (\bibinfo {year} {2016})}\BibitemShut {NoStop}%
\bibitem [{\citenamefont {Porezag}\ and\ \citenamefont
  {Pederson}(1996)}]{PhysRevB.54.7830}%
  \BibitemOpen
  \bibfield  {author} {\bibinfo {author} {\bibfnamefont {D.}~\bibnamefont
  {Porezag}}\ and\ \bibinfo {author} {\bibfnamefont {M.~R.}\ \bibnamefont
  {Pederson}},\ }\bibfield  {title} {\bibinfo {title} {Infrared intensities and
  raman-scattering activities within density-functional theory},\ }\href
  {https://doi.org/10.1103/PhysRevB.54.7830} {\bibfield  {journal} {\bibinfo
  {journal} {Phys. Rev. B}\ }\textbf {\bibinfo {volume} {54}},\ \bibinfo
  {pages} {7830} (\bibinfo {year} {1996})}\BibitemShut {NoStop}%
\bibitem [{\citenamefont {Steele}\ and\ \citenamefont
  {King}(1972)}]{steele1972theory}%
  \BibitemOpen
  \bibfield  {author} {\bibinfo {author} {\bibfnamefont {D.}~\bibnamefont
  {Steele}}\ and\ \bibinfo {author} {\bibfnamefont {W.~T.}\ \bibnamefont
  {King}},\ }\href@noop {} {\bibinfo {title} {Theory of vibrational
  spectroscopy}} (\bibinfo {year} {1972})\BibitemShut {NoStop}%
\bibitem [{\citenamefont {Fonari}\ and\ \citenamefont
  {Stauffer}(2013)}]{vasp_raman_py}%
  \BibitemOpen
  \bibfield  {author} {\bibinfo {author} {\bibfnamefont {A.}~\bibnamefont
  {Fonari}}\ and\ \bibinfo {author} {\bibfnamefont {S.}~\bibnamefont
  {Stauffer}},\ }\href@noop {} {\emph {\bibinfo {title} {vasp\_raman.py}}}\
  (\bibinfo  {publisher} {https://github.com/raman-sc/VASP/},\ \bibinfo {year}
  {2013})\BibitemShut {NoStop}%
\bibitem [{\citenamefont {Skelton}\ \emph {et~al.}(2017)\citenamefont
  {Skelton}, \citenamefont {Burton}, \citenamefont {Jackson}, \citenamefont
  {Oba}, \citenamefont {Parker},\ and\ \citenamefont
  {Walsh}}]{skelton2017lattice}%
  \BibitemOpen
  \bibfield  {author} {\bibinfo {author} {\bibfnamefont {J.~M.}\ \bibnamefont
  {Skelton}}, \bibinfo {author} {\bibfnamefont {L.~A.}\ \bibnamefont {Burton}},
  \bibinfo {author} {\bibfnamefont {A.~J.}\ \bibnamefont {Jackson}}, \bibinfo
  {author} {\bibfnamefont {F.}~\bibnamefont {Oba}}, \bibinfo {author}
  {\bibfnamefont {S.~C.}\ \bibnamefont {Parker}},\ and\ \bibinfo {author}
  {\bibfnamefont {A.}~\bibnamefont {Walsh}},\ }\bibfield  {title} {\bibinfo
  {title} {Lattice dynamics of the tin sulphides sns 2, sns and sn 2 s 3:
  vibrational spectra and thermal transport},\ }\href@noop {} {\bibfield
  {journal} {\bibinfo  {journal} {Physical Chemistry Chemical Physics}\
  }\textbf {\bibinfo {volume} {19}},\ \bibinfo {pages} {12452} (\bibinfo {year}
  {2017})}\BibitemShut {NoStop}%
\bibitem [{\citenamefont {Komornicki}\ and\ \citenamefont
  {McIver~Jr}(1979)}]{komornicki1979efficient}%
  \BibitemOpen
  \bibfield  {author} {\bibinfo {author} {\bibfnamefont {A.}~\bibnamefont
  {Komornicki}}\ and\ \bibinfo {author} {\bibfnamefont {J.~W.}\ \bibnamefont
  {McIver~Jr}},\ }\bibfield  {title} {\bibinfo {title} {An efficient ab initio
  method for computing infrared and raman intensities: application to
  ethylene},\ }\href@noop {} {\bibfield  {journal} {\bibinfo  {journal} {The
  Journal of Chemical Physics}\ }\textbf {\bibinfo {volume} {70}},\ \bibinfo
  {pages} {2014} (\bibinfo {year} {1979})}\BibitemShut {NoStop}%
\bibitem [{\citenamefont {Komornicki}\ and\ \citenamefont
  {Jaffe}(2008)}]{komornicki792}%
  \BibitemOpen
  \bibfield  {author} {\bibinfo {author} {\bibfnamefont {A.}~\bibnamefont
  {Komornicki}}\ and\ \bibinfo {author} {\bibfnamefont {R.~L.}\ \bibnamefont
  {Jaffe}},\ }\bibfield  {title} {\bibinfo {title} {{An ab initio investigation
  of the structure, vibrational frequencies, and intensities of HO2 and
  HOCl}},\ }\href {https://doi.org/10.1063/1.438588} {\bibfield  {journal}
  {\bibinfo  {journal} {The Journal of Chemical Physics}\ }\textbf {\bibinfo
  {volume} {71}},\ \bibinfo {pages} {2150} (\bibinfo {year} {2008})},\ \Eprint
  {https://arxiv.org/abs/https://pubs.aip.org/aip/jcp/article-pdf/71/5/2150/11319209/2150\_1\_online.pdf}
  {https://pubs.aip.org/aip/jcp/article-pdf/71/5/2150/11319209/2150\_1\_online.pdf}
  \BibitemShut {NoStop}%
\bibitem [{\citenamefont {Fredkin}\ \emph {et~al.}(1983)\citenamefont
  {Fredkin}, \citenamefont {Komornicki}, \citenamefont {White},\ and\
  \citenamefont {Wilson}}]{komornicki83}%
  \BibitemOpen
  \bibfield  {author} {\bibinfo {author} {\bibfnamefont {D.~R.}\ \bibnamefont
  {Fredkin}}, \bibinfo {author} {\bibfnamefont {A.}~\bibnamefont {Komornicki}},
  \bibinfo {author} {\bibfnamefont {S.~R.}\ \bibnamefont {White}},\ and\
  \bibinfo {author} {\bibfnamefont {K.~R.}\ \bibnamefont {Wilson}},\ }\bibfield
   {title} {\bibinfo {title} {{Ab initio infrared and Raman spectra}},\ }\href
  {https://doi.org/10.1063/1.444751} {\bibfield  {journal} {\bibinfo  {journal}
  {The Journal of Chemical Physics}\ }\textbf {\bibinfo {volume} {78}},\
  \bibinfo {pages} {7077} (\bibinfo {year} {1983})},\ \Eprint
  {https://arxiv.org/abs/https://pubs.aip.org/aip/jcp/article-pdf/78/12/7077/11188017/7077\_1\_online.pdf}
  {https://pubs.aip.org/aip/jcp/article-pdf/78/12/7077/11188017/7077\_1\_online.pdf}
  \BibitemShut {NoStop}%
\bibitem [{\citenamefont {Giannozzi}\ \emph {et~al.}(2009)\citenamefont
  {Giannozzi}, \citenamefont {Baroni}, \citenamefont {Bonini}, \citenamefont
  {Calandra}, \citenamefont {Car}, \citenamefont {Cavazzoni}, \citenamefont
  {Ceresoli}, \citenamefont {Chiarotti}, \citenamefont {Cococcioni},
  \citenamefont {Dabo} \emph {et~al.}}]{giannozzi2009quantum}%
  \BibitemOpen
  \bibfield  {author} {\bibinfo {author} {\bibfnamefont {P.}~\bibnamefont
  {Giannozzi}}, \bibinfo {author} {\bibfnamefont {S.}~\bibnamefont {Baroni}},
  \bibinfo {author} {\bibfnamefont {N.}~\bibnamefont {Bonini}}, \bibinfo
  {author} {\bibfnamefont {M.}~\bibnamefont {Calandra}}, \bibinfo {author}
  {\bibfnamefont {R.}~\bibnamefont {Car}}, \bibinfo {author} {\bibfnamefont
  {C.}~\bibnamefont {Cavazzoni}}, \bibinfo {author} {\bibfnamefont
  {D.}~\bibnamefont {Ceresoli}}, \bibinfo {author} {\bibfnamefont {G.~L.}\
  \bibnamefont {Chiarotti}}, \bibinfo {author} {\bibfnamefont {M.}~\bibnamefont
  {Cococcioni}}, \bibinfo {author} {\bibfnamefont {I.}~\bibnamefont {Dabo}},
  \emph {et~al.},\ }\bibfield  {title} {\bibinfo {title} {Quantum espresso: a
  modular and open-source software project for quantum simulations of
  materials},\ }\href@noop {} {\bibfield  {journal} {\bibinfo  {journal}
  {Journal of physics: Condensed matter}\ }\textbf {\bibinfo {volume} {21}},\
  \bibinfo {pages} {395502} (\bibinfo {year} {2009})}\BibitemShut {NoStop}%
\bibitem [{\citenamefont {Umari}\ and\ \citenamefont
  {Pasquarello}(2005)}]{umari2005infrared}%
  \BibitemOpen
  \bibfield  {author} {\bibinfo {author} {\bibfnamefont {P.}~\bibnamefont
  {Umari}}\ and\ \bibinfo {author} {\bibfnamefont {A.}~\bibnamefont
  {Pasquarello}},\ }\bibfield  {title} {\bibinfo {title} {Infrared and raman
  spectra of disordered materials from first principles},\ }\href@noop {}
  {\bibfield  {journal} {\bibinfo  {journal} {Diamond and related materials}\
  }\textbf {\bibinfo {volume} {14}},\ \bibinfo {pages} {1255} (\bibinfo {year}
  {2005})}\BibitemShut {NoStop}%
\bibitem [{\citenamefont {Giacomazzi}\ \emph {et~al.}(2005)\citenamefont
  {Giacomazzi}, \citenamefont {Umari},\ and\ \citenamefont
  {Pasquarello}}]{giacomazzi2005medium}%
  \BibitemOpen
  \bibfield  {author} {\bibinfo {author} {\bibfnamefont {L.}~\bibnamefont
  {Giacomazzi}}, \bibinfo {author} {\bibfnamefont {P.}~\bibnamefont {Umari}},\
  and\ \bibinfo {author} {\bibfnamefont {A.}~\bibnamefont {Pasquarello}},\
  }\bibfield  {title} {\bibinfo {title} {Medium-range structural properties of
  vitreous germania obtained through first-principles analysis of vibrational
  spectra},\ }\href@noop {} {\bibfield  {journal} {\bibinfo  {journal}
  {Physical review letters}\ }\textbf {\bibinfo {volume} {95}},\ \bibinfo
  {pages} {075505} (\bibinfo {year} {2005})}\BibitemShut {NoStop}%
\bibitem [{\citenamefont {Giacomazzi}\ and\ \citenamefont
  {Pasquarello}(2007)}]{giacomazzi2007vibrational}%
  \BibitemOpen
  \bibfield  {author} {\bibinfo {author} {\bibfnamefont {L.}~\bibnamefont
  {Giacomazzi}}\ and\ \bibinfo {author} {\bibfnamefont {A.}~\bibnamefont
  {Pasquarello}},\ }\bibfield  {title} {\bibinfo {title} {Vibrational spectra
  of vitreous sio2 and vitreous geo2 from first principles},\ }\href@noop {}
  {\bibfield  {journal} {\bibinfo  {journal} {Journal of Physics: Condensed
  Matter}\ }\textbf {\bibinfo {volume} {19}},\ \bibinfo {pages} {415112}
  (\bibinfo {year} {2007})}\BibitemShut {NoStop}%
\bibitem [{\citenamefont {Giacomazzi}\ \emph {et~al.}(2009)\citenamefont
  {Giacomazzi}, \citenamefont {Umari},\ and\ \citenamefont
  {Pasquarello}}]{giacomazzi2009medium}%
  \BibitemOpen
  \bibfield  {author} {\bibinfo {author} {\bibfnamefont {L.}~\bibnamefont
  {Giacomazzi}}, \bibinfo {author} {\bibfnamefont {P.}~\bibnamefont {Umari}},\
  and\ \bibinfo {author} {\bibfnamefont {A.}~\bibnamefont {Pasquarello}},\
  }\bibfield  {title} {\bibinfo {title} {Medium-range structure of vitreous sio
  2 obtained through first-principles investigation of vibrational spectra},\
  }\href@noop {} {\bibfield  {journal} {\bibinfo  {journal} {Physical Review
  B}\ }\textbf {\bibinfo {volume} {79}},\ \bibinfo {pages} {064202} (\bibinfo
  {year} {2009})}\BibitemShut {NoStop}%
\bibitem [{\citenamefont {Khoo}\ \emph {et~al.}(2010)\citenamefont {Khoo},
  \citenamefont {Zayak}, \citenamefont {Kwak},\ and\ \citenamefont
  {Chelikowsky}}]{khoo2010first}%
  \BibitemOpen
  \bibfield  {author} {\bibinfo {author} {\bibfnamefont {K.}~\bibnamefont
  {Khoo}}, \bibinfo {author} {\bibfnamefont {A.}~\bibnamefont {Zayak}},
  \bibinfo {author} {\bibfnamefont {H.}~\bibnamefont {Kwak}},\ and\ \bibinfo
  {author} {\bibfnamefont {J.~R.}\ \bibnamefont {Chelikowsky}},\ }\bibfield
  {title} {\bibinfo {title} {First-principles study of confinement effects on
  the raman spectra of si nanocrystals},\ }\href@noop {} {\bibfield  {journal}
  {\bibinfo  {journal} {Physical review letters}\ }\textbf {\bibinfo {volume}
  {105}},\ \bibinfo {pages} {115504} (\bibinfo {year} {2010})}\BibitemShut
  {NoStop}%
\bibitem [{\citenamefont {Kilymis}\ \emph {et~al.}(2019)\citenamefont
  {Kilymis}, \citenamefont {Ispas}, \citenamefont {Hehlen}, \citenamefont
  {Peuget},\ and\ \citenamefont {Delaye}}]{kilymis2019vibrational}%
  \BibitemOpen
  \bibfield  {author} {\bibinfo {author} {\bibfnamefont {D.}~\bibnamefont
  {Kilymis}}, \bibinfo {author} {\bibfnamefont {S.}~\bibnamefont {Ispas}},
  \bibinfo {author} {\bibfnamefont {B.}~\bibnamefont {Hehlen}}, \bibinfo
  {author} {\bibfnamefont {S.}~\bibnamefont {Peuget}},\ and\ \bibinfo {author}
  {\bibfnamefont {J.-M.}\ \bibnamefont {Delaye}},\ }\bibfield  {title}
  {\bibinfo {title} {Vibrational properties of sodosilicate glasses from
  first-principles calculations},\ }\href@noop {} {\bibfield  {journal}
  {\bibinfo  {journal} {Physical Review B}\ }\textbf {\bibinfo {volume} {99}},\
  \bibinfo {pages} {054209} (\bibinfo {year} {2019})}\BibitemShut {NoStop}%
\bibitem [{\citenamefont {Giacomazzi}\ \emph {et~al.}(2023)\citenamefont
  {Giacomazzi}, \citenamefont {Shcheblanov}, \citenamefont {Povarnitsyn},
  \citenamefont {Li}, \citenamefont {Mavri{\v{c}}}, \citenamefont
  {Zupan{\v{c}}i{\v{c}}}, \citenamefont {Grdadolnik},\ and\ \citenamefont
  {Pasquarello}}]{giacomazzi2023infrared}%
  \BibitemOpen
  \bibfield  {author} {\bibinfo {author} {\bibfnamefont {L.}~\bibnamefont
  {Giacomazzi}}, \bibinfo {author} {\bibfnamefont {N.~S.}\ \bibnamefont
  {Shcheblanov}}, \bibinfo {author} {\bibfnamefont {M.~E.}\ \bibnamefont
  {Povarnitsyn}}, \bibinfo {author} {\bibfnamefont {Y.}~\bibnamefont {Li}},
  \bibinfo {author} {\bibfnamefont {A.}~\bibnamefont {Mavri{\v{c}}}}, \bibinfo
  {author} {\bibfnamefont {B.}~\bibnamefont {Zupan{\v{c}}i{\v{c}}}}, \bibinfo
  {author} {\bibfnamefont {J.}~\bibnamefont {Grdadolnik}},\ and\ \bibinfo
  {author} {\bibfnamefont {A.}~\bibnamefont {Pasquarello}},\ }\bibfield
  {title} {\bibinfo {title} {Infrared spectra in amorphous alumina: A combined
  ab initio and experimental study},\ }\href@noop {} {\bibfield  {journal}
  {\bibinfo  {journal} {Physical Review Materials}\ }\textbf {\bibinfo {volume}
  {7}},\ \bibinfo {pages} {045604} (\bibinfo {year} {2023})}\BibitemShut
  {NoStop}%
\bibitem [{\citenamefont {Lazzeri}\ and\ \citenamefont
  {Mauri}(2003)}]{rlarge1}%
  \BibitemOpen
  \bibfield  {author} {\bibinfo {author} {\bibfnamefont {M.}~\bibnamefont
  {Lazzeri}}\ and\ \bibinfo {author} {\bibfnamefont {F.}~\bibnamefont
  {Mauri}},\ }\bibfield  {title} {\bibinfo {title} {First-principles
  calculation of vibrational raman spectra in large systems: Signature of small
  rings in crystalline $\textrm{SiO}_ 2$},\ }\href
  {https://doi.org/10.1103/PhysRevLett.90.036401} {\bibfield  {journal}
  {\bibinfo  {journal} {Physical review letters}\ }\textbf {\bibinfo {volume}
  {90}},\ \bibinfo {pages} {036401} (\bibinfo {year} {2003})}\BibitemShut
  {NoStop}%
\bibitem [{\citenamefont {Collins}\ and\ \citenamefont
  {Bettens}(2015)}]{collins2015energy}%
  \BibitemOpen
  \bibfield  {author} {\bibinfo {author} {\bibfnamefont {M.~A.}\ \bibnamefont
  {Collins}}\ and\ \bibinfo {author} {\bibfnamefont {R.~P.}\ \bibnamefont
  {Bettens}},\ }\bibfield  {title} {\bibinfo {title} {Energy-based molecular
  fragmentation methods},\ }\href@noop {} {\bibfield  {journal} {\bibinfo
  {journal} {Chemical reviews}\ }\textbf {\bibinfo {volume} {115}},\ \bibinfo
  {pages} {5607} (\bibinfo {year} {2015})}\BibitemShut {NoStop}%
\bibitem [{\citenamefont {Raghavachari}\ and\ \citenamefont
  {Saha}(2015)}]{raghavachari2015accurate}%
  \BibitemOpen
  \bibfield  {author} {\bibinfo {author} {\bibfnamefont {K.}~\bibnamefont
  {Raghavachari}}\ and\ \bibinfo {author} {\bibfnamefont {A.}~\bibnamefont
  {Saha}},\ }\bibfield  {title} {\bibinfo {title} {Accurate composite and
  fragment-based quantum chemical models for large molecules},\ }\href@noop {}
  {\bibfield  {journal} {\bibinfo  {journal} {Chemical reviews}\ }\textbf
  {\bibinfo {volume} {115}},\ \bibinfo {pages} {5643} (\bibinfo {year}
  {2015})}\BibitemShut {NoStop}%
\bibitem [{\citenamefont {Li}\ \emph {et~al.}(2014)\citenamefont {Li},
  \citenamefont {Li},\ and\ \citenamefont {Ma}}]{doi:10.1021/ar500038z}%
  \BibitemOpen
  \bibfield  {author} {\bibinfo {author} {\bibfnamefont {S.}~\bibnamefont
  {Li}}, \bibinfo {author} {\bibfnamefont {W.}~\bibnamefont {Li}},\ and\
  \bibinfo {author} {\bibfnamefont {J.}~\bibnamefont {Ma}},\ }\bibfield
  {title} {\bibinfo {title} {Generalized energy-based fragmentation approach
  and its applications to macromolecules and molecular aggregates},\ }\href
  {https://doi.org/10.1021/ar500038z} {\bibfield  {journal} {\bibinfo
  {journal} {Accounts of Chemical Research}\ }\textbf {\bibinfo {volume}
  {47}},\ \bibinfo {pages} {2712} (\bibinfo {year} {2014})},\ \bibinfo {note}
  {pMID: 24873495},\ \Eprint
  {https://arxiv.org/abs/https://doi.org/10.1021/ar500038z}
  {https://doi.org/10.1021/ar500038z} \BibitemShut {NoStop}%
\bibitem [{\citenamefont {Khire}\ \emph {et~al.}(2018)\citenamefont {Khire},
  \citenamefont {Sahu},\ and\ \citenamefont {Gadre}}]{Khire}%
  \BibitemOpen
  \bibfield  {author} {\bibinfo {author} {\bibfnamefont {S.}~\bibnamefont
  {Khire}}, \bibinfo {author} {\bibfnamefont {N.}~\bibnamefont {Sahu}},\ and\
  \bibinfo {author} {\bibfnamefont {S.}~\bibnamefont {Gadre}},\ }\bibfield
  {title} {\bibinfo {title} {Harnessing desktop computers for ab initio
  calculation of vibrational ir/raman spectra of large molecules},\ }\href
  {https://doi.org/10.1007/s12039-018-1568-3} {\bibfield  {journal} {\bibinfo
  {journal} {Journal of Chemical Sciences}\ }\textbf {\bibinfo {volume} {130}}
  (\bibinfo {year} {2018})}\BibitemShut {NoStop}%
\bibitem [{\citenamefont {Kresse}\ and\ \citenamefont
  {Furthm\"uller}(1996{\natexlab{a}})}]{PhysRevB.54.11169}%
  \BibitemOpen
  \bibfield  {author} {\bibinfo {author} {\bibfnamefont {G.}~\bibnamefont
  {Kresse}}\ and\ \bibinfo {author} {\bibfnamefont {J.}~\bibnamefont
  {Furthm\"uller}},\ }\bibfield  {title} {\bibinfo {title} {Efficient iterative
  schemes for ab initio total-energy calculations using a plane-wave basis
  set},\ }\href {https://doi.org/10.1103/PhysRevB.54.11169} {\bibfield
  {journal} {\bibinfo  {journal} {Phys. Rev. B}\ }\textbf {\bibinfo {volume}
  {54}},\ \bibinfo {pages} {11169} (\bibinfo {year}
  {1996}{\natexlab{a}})}\BibitemShut {NoStop}%
\bibitem [{\citenamefont {Kresse}\ and\ \citenamefont
  {Furthm\"uller}(1996{\natexlab{b}})}]{CMS_6_15}%
  \BibitemOpen
  \bibfield  {author} {\bibinfo {author} {\bibfnamefont {G.}~\bibnamefont
  {Kresse}}\ and\ \bibinfo {author} {\bibfnamefont {J.}~\bibnamefont
  {Furthm\"uller}},\ }\bibfield  {title} {\bibinfo {title} {Efficiency of
  ab-initio total energy calculations for metals and semiconductors using a
  plane-wave basis set},\ }\href {https://doi.org/10.1016/0927-0256(96)00008-0}
  {\bibfield  {journal} {\bibinfo  {journal} {Comput. Mater. Sci.}\ }\textbf
  {\bibinfo {volume} {6}},\ \bibinfo {pages} {15 } (\bibinfo {year}
  {1996}{\natexlab{b}})}\BibitemShut {NoStop}%
\bibitem [{\citenamefont {Placzek}(1959)}]{placzek1959rayleigh}%
  \BibitemOpen
  \bibfield  {author} {\bibinfo {author} {\bibfnamefont {G.}~\bibnamefont
  {Placzek}},\ }\href@noop {} {\emph {\bibinfo {title} {The rayleigh and raman
  scattering}}},\ Vol.\ \bibinfo {volume} {526}\ (\bibinfo  {publisher}
  {Lawrence Radiation Laboratory},\ \bibinfo {year} {1959})\BibitemShut
  {NoStop}%
\bibitem [{\citenamefont {Mitroy}\ \emph {et~al.}(2010)\citenamefont {Mitroy},
  \citenamefont {Safronova},\ and\ \citenamefont {Clark}}]{Mitroy_2010}%
  \BibitemOpen
  \bibfield  {author} {\bibinfo {author} {\bibfnamefont {J.}~\bibnamefont
  {Mitroy}}, \bibinfo {author} {\bibfnamefont {M.~S.}\ \bibnamefont
  {Safronova}},\ and\ \bibinfo {author} {\bibfnamefont {C.~W.}\ \bibnamefont
  {Clark}},\ }\bibfield  {title} {\bibinfo {title} {Theory and applications of
  atomic and ionic polarizabilities},\ }\href
  {https://doi.org/10.1088/0953-4075/43/20/202001} {\bibfield  {journal}
  {\bibinfo  {journal} {Journal of Physics B: Atomic, Molecular and Optical
  Physics}\ }\textbf {\bibinfo {volume} {43}},\ \bibinfo {pages} {202001}
  (\bibinfo {year} {2010})}\BibitemShut {NoStop}%
\bibitem [{\citenamefont {Project}(2020)}]{osti_1208347}%
  \BibitemOpen
  \bibfield  {author} {\bibinfo {author} {\bibfnamefont {T.~M.}\ \bibnamefont
  {Project}},\ }\bibfield  {title} {\bibinfo {title} {Materials data on geo2 by
  materials project}\ }\href {https://doi.org/10.17188/1208347}
  {10.17188/1208347} (\bibinfo {year} {2020})\BibitemShut {NoStop}%
\bibitem [{\citenamefont {Mernagh}\ and\ \citenamefont
  {Liu}(1997)}]{mernagh1997temperature}%
  \BibitemOpen
  \bibfield  {author} {\bibinfo {author} {\bibfnamefont {T.~P.}\ \bibnamefont
  {Mernagh}}\ and\ \bibinfo {author} {\bibfnamefont {L.-g.}\ \bibnamefont
  {Liu}},\ }\bibfield  {title} {\bibinfo {title} {Temperature dependence of
  raman spectra of the quartz-and rutile-types of $\textrm{GeO}_2$},\
  }\href@noop {} {\bibfield  {journal} {\bibinfo  {journal} {Physics and
  chemistry of minerals}\ }\textbf {\bibinfo {volume} {24}},\ \bibinfo {pages}
  {7} (\bibinfo {year} {1997})}\BibitemShut {NoStop}%
\bibitem [{\citenamefont {Gakiya-Teruya}\ \emph {et~al.}(2021)\citenamefont
  {Gakiya-Teruya}, \citenamefont {Jiang}, \citenamefont {Le}, \citenamefont
  {Ungor}, \citenamefont {Durrani}, \citenamefont {Koptur-Palenchar},
  \citenamefont {Jiang}, \citenamefont {Jiang}, \citenamefont {Meisel},
  \citenamefont {Cheng}, \citenamefont {Zhang}, \citenamefont {Zhang},
  \citenamefont {Rahman}, \citenamefont {Hebard},\ and\ \citenamefont
  {Shatruk}}]{doi:10.1021/jacs.1c04598}%
  \BibitemOpen
  \bibfield  {author} {\bibinfo {author} {\bibfnamefont {M.}~\bibnamefont
  {Gakiya-Teruya}}, \bibinfo {author} {\bibfnamefont {X.}~\bibnamefont
  {Jiang}}, \bibinfo {author} {\bibfnamefont {D.}~\bibnamefont {Le}}, \bibinfo
  {author} {\bibfnamefont {O.}~\bibnamefont {Ungor}}, \bibinfo {author}
  {\bibfnamefont {A.~J.}\ \bibnamefont {Durrani}}, \bibinfo {author}
  {\bibfnamefont {J.~J.}\ \bibnamefont {Koptur-Palenchar}}, \bibinfo {author}
  {\bibfnamefont {J.}~\bibnamefont {Jiang}}, \bibinfo {author} {\bibfnamefont
  {T.}~\bibnamefont {Jiang}}, \bibinfo {author} {\bibfnamefont {M.~W.}\
  \bibnamefont {Meisel}}, \bibinfo {author} {\bibfnamefont {H.-P.}\
  \bibnamefont {Cheng}}, \bibinfo {author} {\bibfnamefont {X.-G.}\ \bibnamefont
  {Zhang}}, \bibinfo {author} {\bibfnamefont {X.-X.}\ \bibnamefont {Zhang}},
  \bibinfo {author} {\bibfnamefont {T.~S.}\ \bibnamefont {Rahman}}, \bibinfo
  {author} {\bibfnamefont {A.~F.}\ \bibnamefont {Hebard}},\ and\ \bibinfo
  {author} {\bibfnamefont {M.}~\bibnamefont {Shatruk}},\ }\bibfield  {title}
  {\bibinfo {title} {Asymmetric design of spin-crossover complexes to increase
  the volatility for surface deposition},\ }\href
  {https://doi.org/10.1021/jacs.1c04598} {\bibfield  {journal} {\bibinfo
  {journal} {Journal of the American Chemical Society}\ }\textbf {\bibinfo
  {volume} {143}},\ \bibinfo {pages} {14563} (\bibinfo {year} {2021})},\
  \bibinfo {note} {pMID: 34472348},\ \Eprint
  {https://arxiv.org/abs/https://doi.org/10.1021/jacs.1c04598}
  {https://doi.org/10.1021/jacs.1c04598} \BibitemShut {NoStop}%
\bibitem [{\citenamefont {Joseph}\ \emph
  {et~al.}(2012{\natexlab{a}})\citenamefont {Joseph}, \citenamefont {Bourson},\
  and\ \citenamefont {Fontana}}]{doi:10.1002/jrs.3142}%
  \BibitemOpen
  \bibfield  {author} {\bibinfo {author} {\bibfnamefont {C.}~\bibnamefont
  {Joseph}}, \bibinfo {author} {\bibfnamefont {P.}~\bibnamefont {Bourson}},\
  and\ \bibinfo {author} {\bibfnamefont {M.~D.}\ \bibnamefont {Fontana}},\
  }\bibfield  {title} {\bibinfo {title} {Amorphous to crystalline
  transformation in $\textrm{Ta}_2\textrm{O}_5$ studied by raman
  spectroscopy},\ }\href {https://doi.org/https://doi.org/10.1002/jrs.3142}
  {\bibfield  {journal} {\bibinfo  {journal} {Journal of Raman Spectroscopy}\
  }\textbf {\bibinfo {volume} {43}},\ \bibinfo {pages} {1146} (\bibinfo {year}
  {2012}{\natexlab{a}})},\ \Eprint
  {https://arxiv.org/abs/https://analyticalsciencejournals.onlinelibrary.wiley.com/\-doi/pdf/10.1002/jrs.3142}
  {https://analyticalsciencejournals.onlinelibrary.wiley.com/\-doi/pdf/10.1002/jrs.3142}
  \BibitemShut {NoStop}%
\bibitem [{\citenamefont {Plimpton}(1995)}]{PLIMPTON19951}%
  \BibitemOpen
  \bibfield  {author} {\bibinfo {author} {\bibfnamefont {S.}~\bibnamefont
  {Plimpton}},\ }\bibfield  {title} {\bibinfo {title} {Fast parallel algorithms
  for short-range molecular dynamics},\ }\href
  {https://doi.org/https://doi.org/10.1006/jcph.1995.1039} {\bibfield
  {journal} {\bibinfo  {journal} {Journal of Computational Physics}\ }\textbf
  {\bibinfo {volume} {117}},\ \bibinfo {pages} {1} (\bibinfo {year}
  {1995})}\BibitemShut {NoStop}%
\bibitem [{\citenamefont {Trinastic}\ \emph {et~al.}(2013)\citenamefont
  {Trinastic}, \citenamefont {Hamdan}, \citenamefont {Wu}, \citenamefont
  {Zhang},\ and\ \citenamefont {Cheng}}]{doi:10.1063/1.4825197}%
  \BibitemOpen
  \bibfield  {author} {\bibinfo {author} {\bibfnamefont {J.~P.}\ \bibnamefont
  {Trinastic}}, \bibinfo {author} {\bibfnamefont {R.}~\bibnamefont {Hamdan}},
  \bibinfo {author} {\bibfnamefont {Y.}~\bibnamefont {Wu}}, \bibinfo {author}
  {\bibfnamefont {L.}~\bibnamefont {Zhang}},\ and\ \bibinfo {author}
  {\bibfnamefont {H.-P.}\ \bibnamefont {Cheng}},\ }\bibfield  {title} {\bibinfo
  {title} {Unified interatomic potential and energy barrier distributions for
  amorphous oxides},\ }\href {https://doi.org/10.1063/1.4825197} {\bibfield
  {journal} {\bibinfo  {journal} {The Journal of Chemical Physics}\ }\textbf
  {\bibinfo {volume} {139}},\ \bibinfo {pages} {154506} (\bibinfo {year}
  {2013})},\ \Eprint {https://arxiv.org/abs/https://doi.org/10.1063/1.4825197}
  {https://doi.org/10.1063/1.4825197} \BibitemShut {NoStop}%
\bibitem [{\citenamefont {Joseph}\ \emph
  {et~al.}(2012{\natexlab{b}})\citenamefont {Joseph}, \citenamefont {Bourson},\
  and\ \citenamefont {Fontana}}]{https://doi.org/10.1002/jrs.3142}%
  \BibitemOpen
  \bibfield  {author} {\bibinfo {author} {\bibfnamefont {C.}~\bibnamefont
  {Joseph}}, \bibinfo {author} {\bibfnamefont {P.}~\bibnamefont {Bourson}},\
  and\ \bibinfo {author} {\bibfnamefont {M.~D.}\ \bibnamefont {Fontana}},\
  }\bibfield  {title} {\bibinfo {title} {Amorphous to crystalline
  transformation in $\textrm{Ta}_2\textrm{O}_5$ studied by raman
  spectroscopy},\ }\href {https://doi.org/https://doi.org/10.1002/jrs.3142}
  {\bibfield  {journal} {\bibinfo  {journal} {Journal of Raman Spectroscopy}\
  }\textbf {\bibinfo {volume} {43}},\ \bibinfo {pages} {1146} (\bibinfo {year}
  {2012}{\natexlab{b}})},\ \Eprint
  {https://arxiv.org/abs/https://analyticalsciencejournals.onlinelibrary.wiley.com/\-doi/pdf/10.1002/jrs.3142}
  {https://analyticalsciencejournals.onlinelibrary.wiley.com/\-doi/pdf/10.1002/jrs.3142}
  \BibitemShut {NoStop}%
\end{thebibliography}%

\end{document}